# Polar compensation at the surface of SrTiO$_3$ (111)

M. Saghayezhian, Lina Chen, Gaomin Wang, Hangwen Guo, E.W. Plummer, and Jiandi Zhang*

*Department of Physics and Astronomy, Louisiana State University, Baton Rouge, Louisiana 70803, USA*

**Abstract**

We have systematically investigated the annealing effect on the structure and composition of the polar surface of SrTiO$_3$ (111), starting with an *ex-situ* chemical etch. The relative surface concentration between Ti and Sr strongly depends on both the annealing temperature and the oxygen processing. There is a critical annealing temperature at which the maximum concentration ratio of Ti to Sr is achieved, while still maintaining a (1×1) surface structure. We demonstrate that with proper processing it is possible to avoid surface reconstruction over a broad temperature range. Our results provide an optimal temperature window for epitaxial film growth.

*email: jiandiz@lsu.edu

**Introduction**

Emergence of novel properties such as superconductivity [1], high-mobility 2D electron gas [2] and topological phases [3-6] at the interface of metal-oxide materials have demonstrated that heterostructure engineering can be an elegant way of fabricating artificial structures displaying new functionalities. Interfaces in the (111) direction of perovskite structures have become a subject of intense study due to the predictions of topological insulator phases and the ability to make artificial double perovskites [7-9]. Recently, unusual electronic and magnetic properties [10-12] have shown the importance of interfaces in the (111) direction. However, processing methods for ideal or well-defined (111) surfaces in film growth are still undetermined [13,14].

According to Tasker's classification [15-17] there are three types of surfaces of ionic crystals: Type I consists of neutral planes. Type II is formed by charged planes which are stacked in a way that the net dipole moment is zero. Type III is constructed by the stacking of charged planes in a way that there is a net dipole moment perpendicular to the planes. Ternary (111) oxide surfaces are classified as type III surfaces which are called polar surfaces. In this type of surface, due to the presence of net dipole moment, the electrostatic surface energy diverges. This is the reason for the instability of type III surfaces which is sometimes referred to as "polar catastrophe". In processing, a polar surface is likely to reconstruct itself to minimize its energy by: structural reconstruction, electronic reconfiguration, changes in surface stoichiometry, faceting, as well as adsorbing atoms/molecules from the environment [17-19]. For oxide film epitaxial growth on such polar surface, a pristine surface without reconstruction is desired in order to reduce the structural and/or chemical complexity at interface.

The (111) direction of SrTiO$_3$ (STO) is formed by stacking Ti$^{4+}$ and [SrO$_3$]$^{4-}$ layers with an inter-planar distance of 2.25 Å, as shown in Fig. 1. There are two common methods to prepare

STO (111). One is by using *in-situ* cycles of ion sputtering and annealing. The resultant surface prepared by this method can be free of contamination, but also exhibit serious structural reconstructions with different chemical composition compared to the bulk. Surface reconstruction, including (n×n) (n=4,5,6) [20,21], (√5×√5)R26.6° and (√13×√13)R33.7° [18], have been observed depending on the details of processing procedure. The surfaces with these reconstructions are not favorable for oxide epitaxy, yielding an interface with complicated structure and chemical intermixing. The other common method to prepare STO (111) surface, e.g., a Ti-terminated surface, is through *ex-situ* selective chemical etching of the SrO layer followed by thermal annealing [22-24]. Although the latter method produces an atomically flat substrate, the epitaxial growth of materials on STO (111) is still complicated, since interface intermixing can occur [13,14,25]. Surface contamination is also an unresolved/unavoidable issue due to *ex-situ* sample preparation. The surface can also reconstruct, depending on processing conditions. Therefore, it is important to determine the structure, chemical composition, and electronic evolution of this surface with different processing procedures, at elevated temperatures that are appropriate for film growth. Understanding how the surface evolves at the desired growth conditions allows one to optimize the growth temperature accordingly.

In this work it is demonstrated that by controlling the annealing temperature, one can remove adsorbates/contaminations, maximize Ti-termination of the surface, compensate the surface polarity, while maintain a (1×1) structure. The results provide an optimal temperature range for oxide film epitaxy.

**Experimental Methods**

In this study, we used Nb-doped SrTiO$_3$ (111). The samples were 0.1% Nb:STO (111), obtained from Crystec GmbH. The samples were cleaned with acetone and ethyl alcohol, then soaked in sonicated deionized (DI)-water with a pH of 5.5 at room temperature for 4 minutes. This was followed by soaking in Buffered HF with the pH of 4.5 for 30 seconds. Finally, samples were annealed at 930 °C for 150 minutes in oxygen. The DI-water etching time and its temperature are important, because etching for too long will cause surface roughness, and high temperature etching will result in faceting after annealing. After *ex-situ* treatment, we post-annealed the samples *in-situ* to remove any unwanted contamination and performed ARXPS. The two sets of samples were annealed for 30 minutes, one in vacuum and the other in an O$_2$/O$_3$ mixture, and then transferred *in-situ* to a characterization chamber for ARXPS measurements.

Angle resolved X-ray photoemission spectroscopy (ARXPS) was utilized to determine the chemical composition near the surface. The core level spectra of O 1s, Ti 2p and Sr 3p were measured using a monochromated Al K$_\alpha$ x-ray source and PHOIBOS 150 energy analyzer, both from SPECS. The energy analyzer was calibrated with the core level of single crystalline gold (Au 4f$_{7/2}$ peak). The depth profile of the components can be extracted from the ARXPS data knowing the relative photoionization cross-sections and the mean free paths [29]. XPS was also used to record contamination of the surface as a function of processing procedures.

Morphology of the surface was measured using an Omicron variable temperature scanning tunneling microscopy (STM), mounted on the main vacuum chamber. We do not have STM image with atomic resolution, so the details of atomic and or electronic reconstruction could not be studied. The lattice symmetry of the surface was determined by the image of low energy electron diffraction (LEED) as a function of the processing. We used an Omicron LEED system, mounted in the main chamber, and the LEED patterns were collected by a CCD camera.

**Results and Discussions**

The morphology and surface structure of the ex-situ prepared sample was investigated in the main UHV chamber using STM and LEED. Figure 2(a) and (b) show a STM image and the line profile, revealing step-terrace morphology with a step height that corresponds to the distance between two Ti layers (Fig. 1). The terrace width depends on the miscut angle, which in our case is ~ 65 Å. The lattice symmetry at the surface is displayed in the LEED pattern shown in Fig. 2(c) & (d), displaying a (1×1) symmetry at $T = 25$ °C and 600 °C, respectively. There is no surface reconstruction observed at higher temperatures (up to 740 °C), although it is difficult to record LEED patterns at this temperature due to radiation from the hot sample. There is no reconstruction to avoid any "polar catastrophe". As previously stated in the introduction, other studies have reported surface reconstruction on the STO (111) surface [18-20,26] after *in-situ* treatments using various sequences of sputtering and annealing. Also, most of the observed reconstructions appear after annealing at very high temperature. For example, 3×3 and 2×2 reconstructions appear after annealing at 1150 °C and 1280 °C, respectively [27].

Chemical composition of the surface as a function of the processing and temperature was determined using ARXPS. Figure 3 summarizes the ARXPS results of Sr and Ti core level spectra, taken at room temperature, for the surface of SrTiO (111), annealed *in-situ* in both vacuum and oxygen at different temperatures. Figure 3(a) and 3(b) show the evolution of the Sr 3p and Ti 2p spectra as a function of annealing (in vacuum) temperature for normal emission ($\theta = 0°$) and emission angle $\theta = 75°$, respectively. Since $\theta = 75°$ data is more surface sensitive, it is easy to see the presence of C contamination, which goes away with high temperature (650 °C) annealing. Using the Tanuma et al. method [28], we calculated the mean free path of Ti 2p (21.1 Å) and Sr 2p (24.3 Å), which illustrates that the average escape depth of electrons at $\theta = 80°$ is about 3.6 Å, such that only the surface layer is probed at high emission angles.

Using sensitivity factors for Ti and Sr, which depend on the photoionization cross section [28] and mean free path of the photoelectron [29], the relative change of the ratio between Ti and Sr core level intensities as a function of emission angle can be converted into the Ti vs. Sr concentration as a function of depth. Figures 3(b) and 3(d) display the measured intensity ratio ($I_{Ti\ 2p}/I_{Sr\ 3p}$) as a function of emission angle, for annealing in vacuum and in oxygen/ozone, respectively. Both procedures of *in-situ* annealing produce a surface that is Ti rich, but there is more Ti present at the vacuum annealed surface, when compared to the oxygen annealed surface. Inspection of the data reveals that the Ti/Sr surface concentration is non-monotonic with annealing temperature, peaking at ~ 660 °C. The insets in Figs. 3(c) and 3(d) show the percentage of the surface that is Ti terminated, assuming that in the bulk the ratio is 1. For vacuum annealing the maximum Ti concentration occurs at ~ 660 °C with a value of 90 % ±10%, i.e. Ti terminated. But for $O_2/O_3$ annealing, the Ti-termination percentage has fallen to ~80% at the same temperature. The conclusion is that *ex-situ* chemical etching followed by vacuum annealing successfully removes the $SrO_3$ layer and creates the best Ti-terminated surface at $T_c$ ~ 660 °C. The introduction of oxygen during annealing reduces the degree of Ti surface termination.

Considerable amount of Carbon (C) contamination is present on the surface before the *in-situ* processing, but disappears at high annealing temperatures. As shown in Fig. 3(b), an appreciable C-core peak appears at ~ 286.2 eV in the spectra and gradually diminishes with increasing annealing temperature. C contamination is commonly observed on STO surfaces before sputtering [30], and could stabilize the polar surface of STO (111). The C 1s signal shown in Fig. 3(b) is ~ 40% of the O 1s signal. Given that the surface is only ~60% Ti [inset Fig. 3(c)] and that the photoionization cross sections are nearly identical, there is a large fraction of a monolayer of C on the surface. The form of this carbon layer can be Ti carbide, graphite, or C bonded to O. The binding energy (286.2 eV) of observed C 1s [Fig. 3(b)] is appreciably higher than the values reported for $C - Ti$ (281.6 eV) or for $C - C$ bond (284.8 eV), but exactly the one for $CO - Ti$ bond [31-33]. This indicates that C forms CO with oxygen and bonds to surface Ti.

Figure 4 shows the $C_{1s}$ intensity as a function of temperature for both vacuum and $O_2/O_3$ annealing. C leaves the surface as annealing temperature increases. As shown in the inset of Fig. 4, the ratio of O 1s to Ti 2p decreases (Ti at surface increases) with increasing temperature, but the ratio of O 1s to Sr 3p remains constant, which means that the carbon leaves while bonded to an oxygen atom. Above 660 °C, the surface is already free from C, the Sr starts to segregate to the surface and the surface starts to lose oxygen, therefore the ratio of O 1s to Sr 3p drops. At high temperature (above ~ 600 °C) the ratio of O 1s to Sr 3p remains constant because, as the C layer is removed, the underlying $SrO_3$ layer is more exposed.

In order to mimic the surface conditions for thin film growth, we carried out ARXPS measurements with the sample held at elevated temperatures. The intensity ratio ($I_{Ti\ 2p}/I_{Sr\ 3p}$), measured at elevated temperature (Fig. 5) displays the same general behavior seen in Fig. 3(c), the Ti/Sr ratio increases up to ~ 660 °C. The binding energies of the Ti 2p, Sr 3p and O 1s core levels also show a temperature dependent shift that looks similar to the intensity ratio. The shift ΔE of the binding energies is basically constant up to ~ 500 °C and then drops by about 1 eV. This shift is not a chemical shift and therefore must be a final state effect. It has been observed using ARPES [34,35] that upon annealing, occupancy at the Fermi surface appears and the surfaced becomes metallic. It is speculated that the creation of oxygen vacancies upon annealing generates carriers to populate the surface state and form a Fermi surface [36]. Semi-empirical Hartree-Fock calculations by Pojani *et al.* [37,38] support this ARPES observation. It is important to note that the drop in the binding energy coincides with the sudden increase of Ti concentration on the surface, thus indicating a correlation of surface metallicity and Ti concentration. During the ARXPS measurements, even at elevated temperatures, no sign of Nb was observed. This could be due to the very small Nb concentration of (only ~ 0.1 %).

Interestingly, the onset of the rising ratio between Ti 2p and Sr 3p, the drop in $\Delta E_b$ and the disappearance in the carbon core level concur around the same temperature. Therefore, it is clear that the STO (111) polar surface takes advantage of contamination as a passivation mechanism, electronic reconfiguration as a screening mechanism, and surface intermixing as the surface reconstruction mechanism, all to minimize the surface energy and avoid polar catastrophe. The mechanisms to avoid "polar catastrophe" and produce polar compensation at the surface are summarized in Fig.5b. First, contamination is a mechanism to avoid polar catastrophe at surface temperatures below 550 °C. Then, electronic reconfiguration takes over as the mechanism in the surface temperature range of 550-660 °C. Finally, as surface temperatures go beyond 660 °C, structural reconstruction, including changes in surface stoichiometry starts to recover the charge neutrality of the surface. Therefore we propose the best temperature window for thin film growth is between 600 °C and 660 °C, in which the surface is free from contamination and is maximally Ti-terminated.

**Conclusion**

The polar surface of STO (111) has been studied for various annealing temperatures. We found that annealing in vacuum and $O_2/O_3$ have different effects on the surface. We observed that annealing in vacuum makes the surface more Ti-rich compared with annealing in oxygen. Prior to annealing, the surface has carbon contamination, which is removed by annealing in vacuum or $O_2/O_3$. The results define a critical annealing temperature at which the maximum surface concentration ratio of Ti to Sr is achieved in a way that the surface is mainly terminated with a Ti-layer, while still maintaining a (1×1) surface structure. We show that with proper processing it is possible to avoid surface reconstruction over a broad temperature range. Our results provide an optimal temperature range for epitaxial growth on this polar surface.

*Acknowledgments*. This work was supported by U.S. DOE under Grant No. DOE DE-SC0002136. G.W. was supported by NSF EPSCoR LA-SiGMA project under award # EPS-1003897. We thank David Howe for critically reading the manuscript.

**References:**

[1] N. Reyren *et al.*, Science **317**, 1196 (2007).
[2] A. Ohtomo and H. Hwang, Nature **427**, 423 (2004).
[3] Y. Wang, Z. Wang, Z. Fang, and X. Dai, arXiv preprint arXiv:1409.6797 (2014).
[4] D. Doennig, W. E. Pickett, and R. Pentcheva, Physical Review B **89**, 121110 (2014).
[5] A. Rüegg, C. Mitra, A. A. Demkov, and G. A. Fiete, Physical Review B **88**, 115146 (2013).
[6] D. Xiao, W. Zhu, Y. Ran, N. Nagaosa, and S. Okamoto, Nat Commun **2**, 596 (2011).
[7] K. Ueda, H. Tabata, and T. Kawai, Science **280**, 1064 (1998).
[8] O. Erten, O. N. Meetei, A. Mukherjee, M. Randeria, N. Trivedi, and P. Woodward, Physical Review Letters **107**, 257201 (2011).
[9] K. I. Kobayashi, T. Kimura, H. Sawada, K. Terakura, and Y. Tokura, Nature **395**, 677 (1998).
[10] B. Lee, O.-U. Kwon, R. H. Shin, W. Jo, and C. U. Jung, Nanoscale research letters **9**, 1 (2014).
[11] B. Kim and B. Min, Physical Review B **89**, 195411 (2014).
[12] R. Sachs, Z. Lin, and J. Shi, Scientific reports **4** (2014).
[13] J. Blok, X. Wan, G. Koster, D. Blank, and G. Rijnders, Applied Physics Letters **99**, 151917 (2011).
[14] S. Middey, P. Rivero, D. Meyers, M. Kareev, X. Liu, Y. Cao, J. Freeland, S. Barraza-Lopez, and J. Chakhalian, Scientific reports **4** (2014).
[15] P. W. Tasker, Journal of Physics C: Solid State Physics **12**, 4977 (1979).
[16] G. Jacek, F. Fabio, and N. Claudine, Reports on Progress in Physics **71**, 016501 (2008).
[17] C. Noguera, Journal of Physics: Condensed Matter **12**, R367 (2000).
[18] B. C. Russell and M. R. Castell, Physical Review B **75**, 155433 (2007).
[19] B. C. Russell and M. R. Castell, The Journal of Physical Chemistry C **112**, 6538 (2008).
[20] A. N. Chiaramonti, C. H. Lanier, L. D. Marks, and P. C. Stair, Surface Science **602**, 3018 (2008).
[21] H. Tanaka and T. Kawai, Surface Science **365**, 437 (1996).
[22] J. Connell, B. Isaac, G. Ekanayake, D. Strachan, and S. Seo, Applied Physics Letters **101**, 251607 (2012).
[23] J. Chang, Y.-S. Park, and S.-K. Kim, Applied Physics Letters **92**, 152910 (2008).
[24] A. Biswas, P. Rossen, C.-H. Yang, W. Siemons, M.-H. Jung, I. Yang, R. Ramesh, and Y. Jeong, Applied Physics Letters **98**, 051904 (2011).
[25] J. Chang, Y.-S. Park, J.-W. Lee, and S.-K. Kim, Journal of crystal growth **311**, 3771 (2009).
[26] J. Feng, X. Zhu, and J. Guo, Surface Science **614**, 38 (2013).
[27] L. D. Marks, A. N. Chiaramonti, S. U. Rahman, and M. R. Castell, Physical Review Letters **114**, 226101 (2015).
[28] S. Tanuma, C. J. Powell, and D. R. Penn, Surface and Interface Analysis **21**, 165 (1994).
[29] M. P. Seah, I. S. Gilmore, and S. J. Spencer, Journal of Electron Spectroscopy and Related Phenomena **120**, 93 (2001).
[30] G. M. Vanacore, L. F. Zagonel, and N. Barrett, Surface Science **604**, 1674 (2010).
[31] A. V. Shchukarev and D. V. Korolkov, Central European Journal of Chemistry **2**, 347.
[32] E. Lewin *et al.*, Surface and Coatings Technology **202**, 3563 (2008).
[33] M. K. Rajumon, M. S. Hegde, and C. N. R. Rao, Catalysis Letters **1**, 351.
[34] N. Plumb *et al.*, Physical review letters **113**, 086801 (2014).
[35] T. C. Rödel *et al.*, Physical Review Applied **1**, 051002 (2014).
[36] S. McKeown Walker *et al.*, Physical Review Letters **113**, 177601 (2014).
[37] A. Pojani, F. Finocchi, and C. Noguera, Surface Science **442**, 179 (1999).
[38] A. Pojani, F. Finocchi, and C. Noguera, Applied Surface Science **142**, 177 (1999).

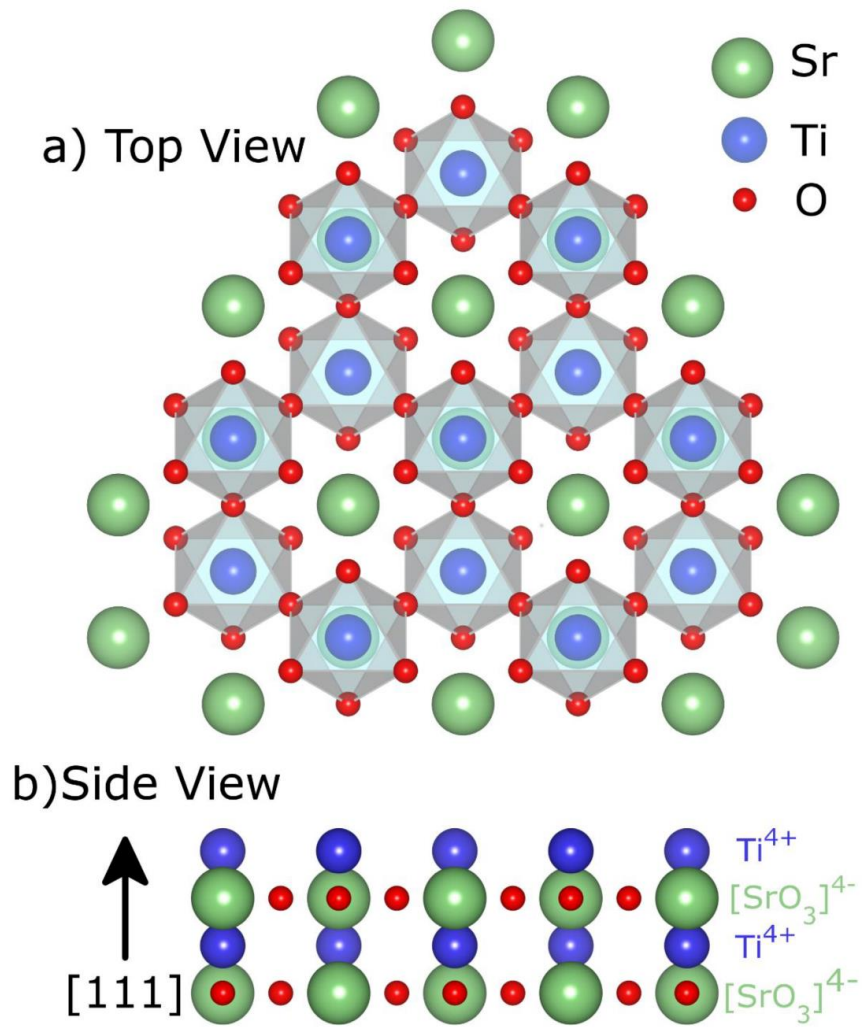

**Fig. 1** (a) Top view and (b) side view of SrTiO$_3$ (111), which is composed of the stacking sequence of oppositely charged [SrO$_3$]$^{4-}$ and Ti$^{4+}$ atomic layers, resulting in a divergent surface energy.

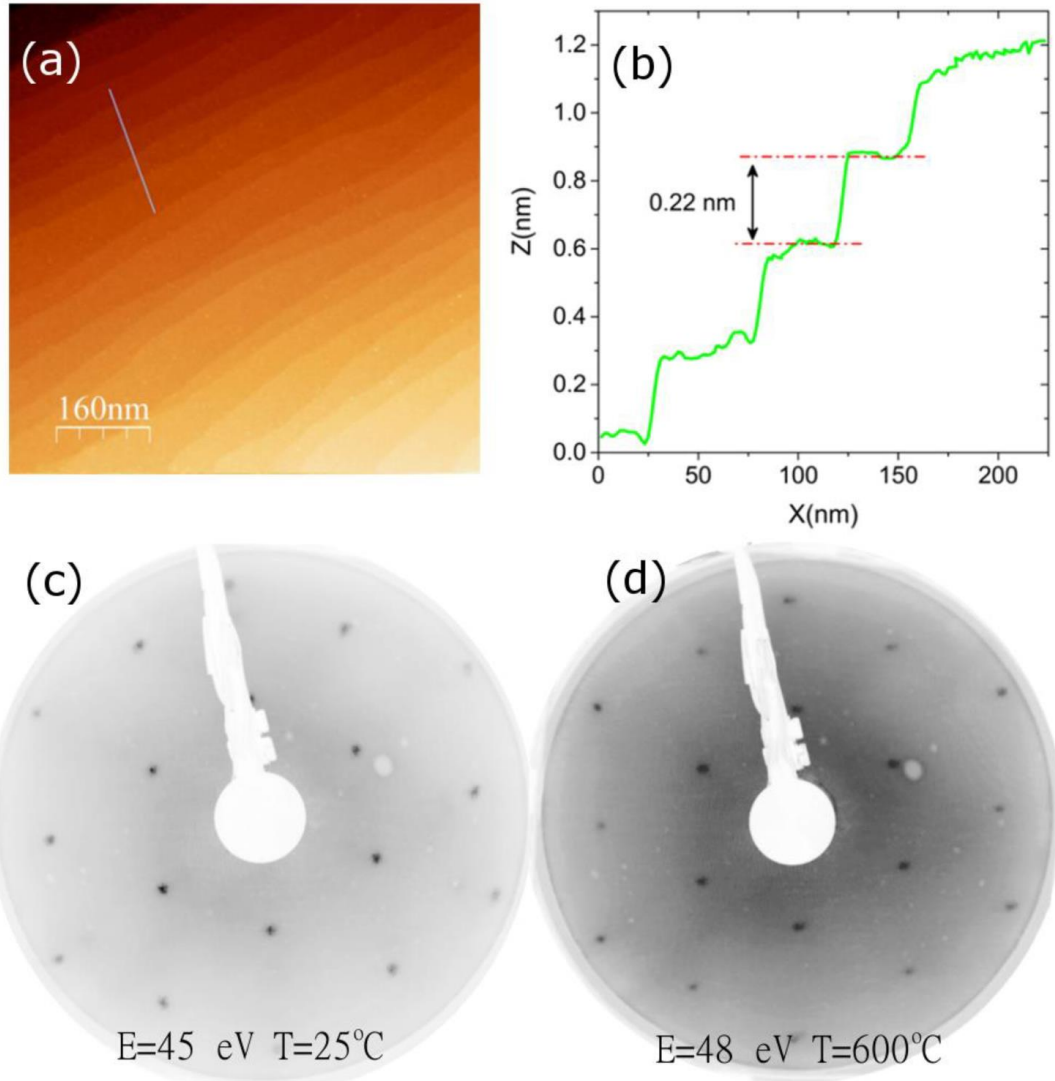

**Fig. 2** (a) STM image of SrTiO$_3$ (111) surface taken at room temperature after preparation and (b) the line profile across the steps marked by the line in (a). (c) and (d) display the LEED $p(1\times1)$ pattern of STO (111) taken at room tempearture and 600 °C, respectively.

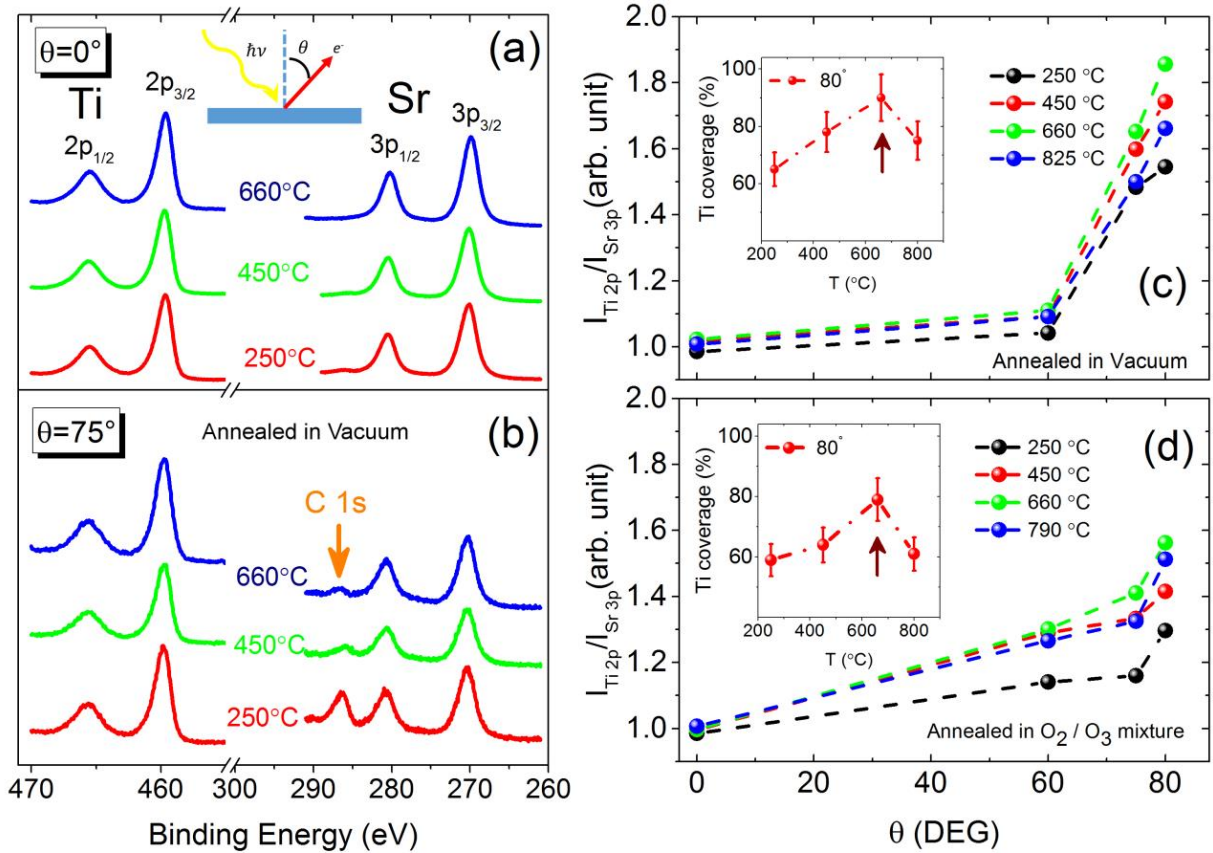

**Fig. 3** ARXPS spectra measured at room temperature for Ti 2p and Sr 3p core levels of SrTiO$_3$ (111) surface with photoelectron emission anlge (a) $\theta = 0°$ and (b) $\theta = 75°$ for different annealing temperatures in vacuum. The inset shows the schematic setup of ARXPS. The arrow points to carbon peak that disapears as the temperature increases. The Ti 2p/Sr 3p intensity ratio as a function of emission angle for the surface annealed at different temperatures (c) without and (d) with oxygen/ozone is plotted. The insets show the ratio as a function of temperature at $\theta = 80°$.

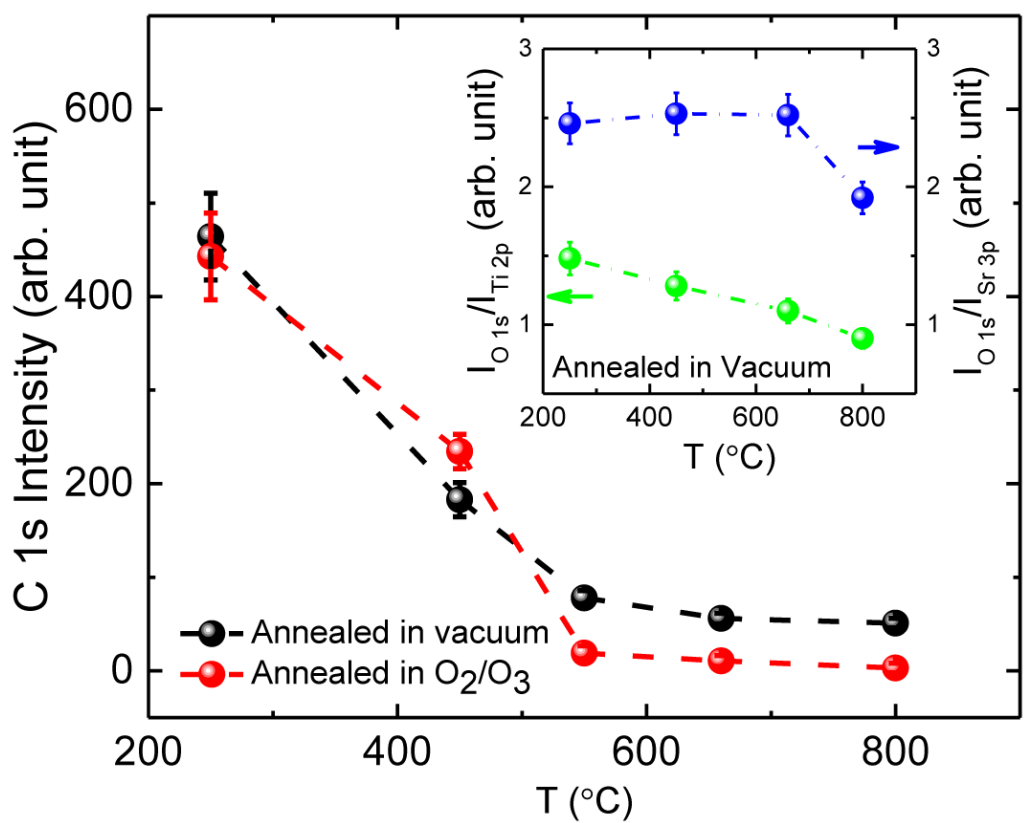

**Fig. 4** The change in carbon intensity as a function of annealing in vacuum and oxygen mixture. The inset shows the change of $O_{1s}/Ti_{2p}$ intensity and $O_{1s}/Sr_{3p}$ intensity annealed in vacuum as a function of temperature. The measurements were performed at room tempreture.

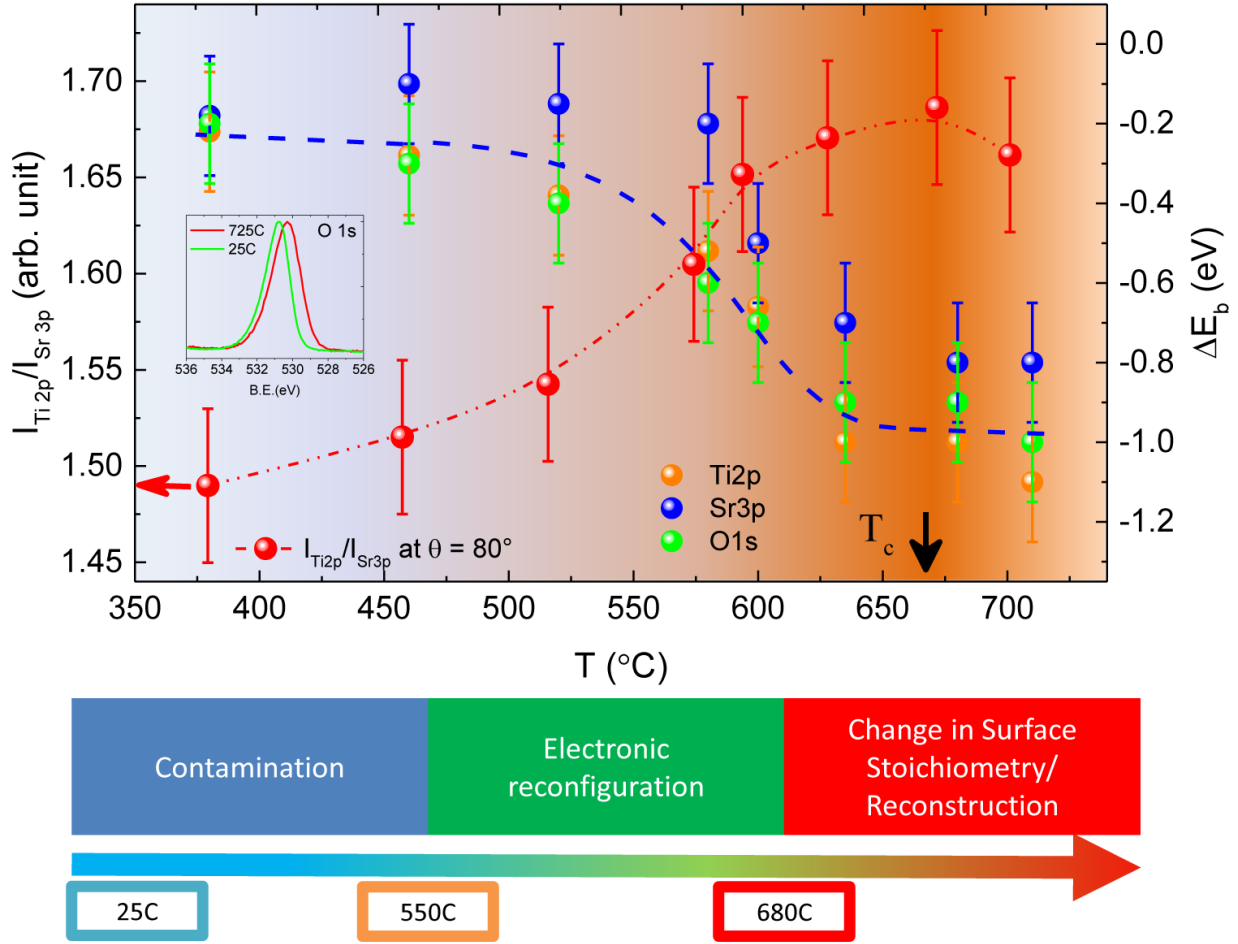

**Fig. 5** (*top*) The temperature-dependence of the binding energy ($\Delta E_b$) for core levels of Ti 2p, Sr 3p and O 1s and the relative intensity ratio of Ti 2p and Sr 3p determined from ARXPS spectra for SrTiO$_3$ (111) surface. The data was taken at photoelectron emission anlge $\theta = 80°$ and elevated sample temperature. The inset shows the binding energy shift of O 1s to lower binding energy at 25 ℃ and 725 ℃ due to final state effect. (*bottom*) The schematic order of different mechanisms to avoid polar catastrophe as a function of temperature.